\documentclass[aps,prb,twocolumn,showpacs,superscriptaddress,floatfix,nofootinbib]{revtex4}
\usepackage{amsfonts}
\usepackage{graphicx}
\usepackage{amsmath}
\usepackage{times}
\usepackage{amssymb}
\usepackage{color}
\usepackage[colorlinks,bookmarks=false,citecolor=blue,linkcolor=red,urlcolor=blue]{hyperref}

\begin{document}

\title{
Tensor network study of the Shastry-Sutherland model in zero magnetic field
}

\author{Philippe Corboz}
\affiliation{Theoretische Physik, ETH Z\"urich, CH-8093 Z\"urich, Switzerland}
\author{Fr\'ed\'eric Mila}
\affiliation{Institut de th\'eorie des ph\'enom\`enes physiques, \'Ecole Polytechnique F\'ed\'erale de Lausanne (EPFL), CH-1015 Lausanne, Switzerland}

\date{\today}

\begin{abstract}
We simulate the Shastry-Sutherland model in two dimensions by means of infinite projected entangled-pair states (iPEPS) - a variational tensor network method where the accuracy can be systematically controlled by the so-called bond dimension. Besides the well established dimer and N\'eel phase iPEPS confirms the presence of an intermediate phase with plaquette long-range order, and we  determine its phase boundaries with high accuracy. The first order phase transition for $J=0.675(2)$ between dimer and plaquette phase is compatible with previous series expansion results. iPEPS predicts a weak first-order phase transition between plaquette and N\'eel phase occurring for $J=0.765(15)$. We do not find a stable intermediate columnar-dimer phase, even when we bias the state towards this order.
\end{abstract}

\maketitle

\section{Introduction}
The Shastry-Sutherland model~\cite{Shastry81} (SSM) has been subject of many theoretical studies over the past three decades. Particularly after the discovery of the two-dimensional spin-gap material SrCu$_2$(BO$_3$)$_2$,~\cite{Kageyama99, Miyahara03} which can  effectively be described by the SSM,~\cite{Miyahara99} theoretical efforts to determine its phase diagram have been intensified. Since the model is frustrated, accurate Quantum Monte Carlo simulations are lacking due to the negative sign problem, and various other methods have led to conflicting conclusions so far.

\begin{figure}
\begin{center}
\includegraphics[width=8.5cm]{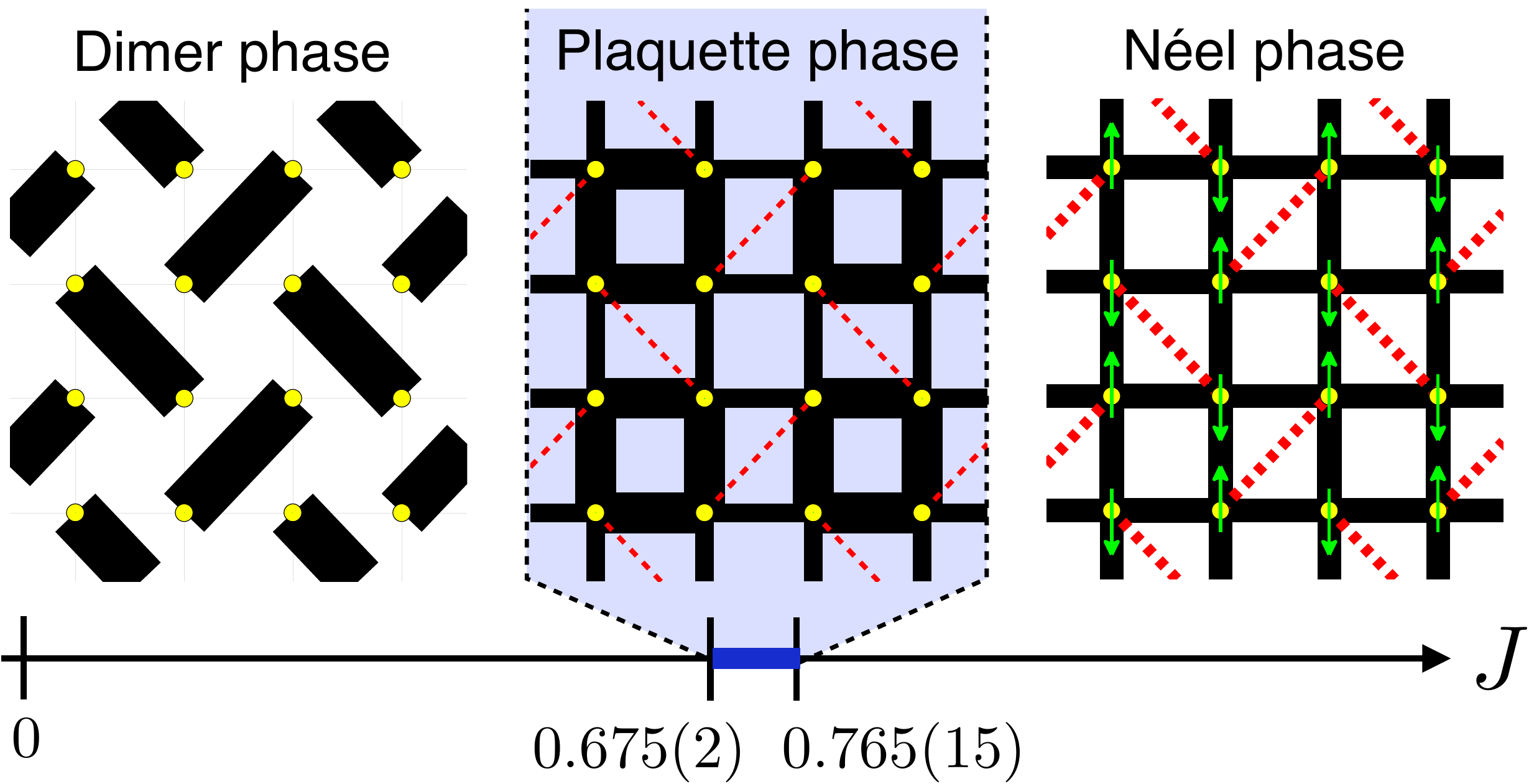}
\caption{(Color online) The phase diagram of the Shastry-Sutherland model as a function of nearest-neighbor coupling $J$ ($J'=1$), obtained with iPEPS. The width of a bond is proportional to the magnitude of the bond energy, where full (dashed) lines correspond to negative (positive) energies. The arrows in the right panel illustrate the N\'eel order. In between the well established Dimer and N\'eel phase we find a phase with plaquette long-range order. }
\label{fig:PD}
\end{center}
\end{figure}

The SSM is given by the following Hamiltonian
\begin{equation}
\label{eqn:H}
\hat H = J \sum_{\langle i,j \rangle} S_i \cdot S_j + J' \sum_{\langle\langle i,j \rangle\rangle_\text{dimer} } S_i \cdot S_j
\end{equation}
where $S_i$ are spin-1/2 operators. The first sum goes over nearest-neighbors on a square lattice, and the second sum over next-nearest neighbor sites  on orthogonal dimers following the pattern shown Fig.~\ref{fig:PD}. In this work we set $J'=1$ and study the phase diagram as a function of $J$. For $J=0$ the model reduces to a Hamiltonian of decoupled dimers with a ground state energy given by a product of singlets with an energy of $-3/4$ per dimer. This state remains the \emph{exact} ground state also for finite $J$ up to a certain value $J_{c1}$.~\cite{Shastry81} The other limit $J\rightarrow \infty$ (or $J'=0$) corresponds to the Heisenberg model, where the ground state exhibits N\'eel order.

One of the first studies based on Schwinger boson mean-field theory predicted an intermediate helical phase between the dimer and the N\'eel ordered phase.~\cite{Albrecht96} Other early works suggested a direct transition between the two phases without any intermediate phase.~\cite{Miyahara99,Zheng99,Muller00}

A plaquette phase as an intermediate phase was first found in the series expansion study by Koga and Kawakami.~\cite{Koga00} They predicted a first order transition between dimer and plaquette phase for $J_{c1}=0.677(2)$, and a second order phase transition between plaquette and N\'eel phase for $J_{c2}=0.86(1)$. This phase has been confirmed in the series expansion study in Ref.~\onlinecite{Takushima01}, where it was shown that the plaquette phase is adiabatically connected to the ground state of the 1/5-depleted square lattice model. A plaquette phase adjacent to the N\'eel phase has also been found in the theoretical study in Ref.~\onlinecite{Chung01} based on a $1/N$ expansion.
Further support for a plaquette state was provided by exact diagonalization and a combination of dimer and quadrumer-boson methods,\cite{Laeuchli02} where characteristic features of the intermediate plaquette phase have been found for $0.678<J<0.702$.

On the other hand, the series expansion study in Ref.~\onlinecite{Zheng01} challenged the prediction of a plaquette state as intermediate phase, and proposed a columnar-dimer ordered state as another possible candidate. However, since none of the proposed non-magnetic states has an energy that is clearly lower than the N\'eel energy (in the relevant coupling regime) they were not able to make a final conclusion regarding the nature of the intermediate phase.

In this work we show that the intermediate phase has indeed plaquette long-range order, and we determine the phase boundaries with a higher accuracy than in previous studies. Our results, summarized in Fig.~\ref{fig:PD}, have been obtained by means of infinite projected entangled-pair states (iPEPS), which is a variational ansatz where the wave function is represented by a \emph{tensor network}.~\cite{verstraete2004,jordan2008} This approach can be seen as a two-dimensional generalization of matrix-product states (MPS) - the underlying variational ansatz of the famous density-matrix renormalization group (DMRG) method.~\cite{white1992} The accuracy of the ansatz can be systematically controlled by the so-called bond dimension $D$. A similar approach has already been successfully used for the study of various bosonic, frustrated spin- and fermionic models (see e.g. Refs.~\onlinecite{jordan2009, corboz2011, bauer2011-su3, Gu11, Wang11, Zhao12, Corboz12_su4}). We note that a finite PEPS has been used in a previous study of the SSM,\cite{Isacsson06} however, only with a small bond dimension $D=2$, and no intermediate phase has been found.

Besides the phase diagram of the SSM, one of the goals of this paper is to provide further benchmark data which demonstrate the performance and usefulness of iPEPS. We present results obtained with different simulation setups which are explicitly biased towards certain orders. However, despite the bias, all simulation setups lead to consistent results in the large $D$ limit. For example, we put a  bias towards columnar dimer order, in which case for small $D$ the dimer order is clearly reproduced. However, it vanishes for large $D$ which shows that  the dimer order is not stable. 
Finally, we present and test a scheme to treat next-nearest neighbor interactions in a more accurate way than in a previous study.~\cite{corboz2010-nn}

The paper is organized as follows: In Sec.~\ref{sec:method} we provide a brief introduction to the iPEPS method and explain the different simulation setups used in this work. In Sec.~\ref{sec:results} we present our simulation results, first for values of $J$ deep in the individual phases, followed by a detailed study of the phase transitions. Finally, in Sec.~\ref{sec:conclusion} we summarize our findings.  In appendix~\ref{sec:nnscheme} the scheme to treat next-nearest neighbor interactions in iPEPS is explained. 

\section{Method}
\label{sec:method}
\subsection{Infinite projected entangled-pair states}
In this section we provide a short overview of iPEPS. For a more detailed introduction to iPEPS and tensor networks in general we refer to Refs.~\onlinecite{Verstraete08,schollwoeck2011,jordan2008,corboz2010}.

\begin{figure}
\begin{center}
\includegraphics[width=8.5cm]{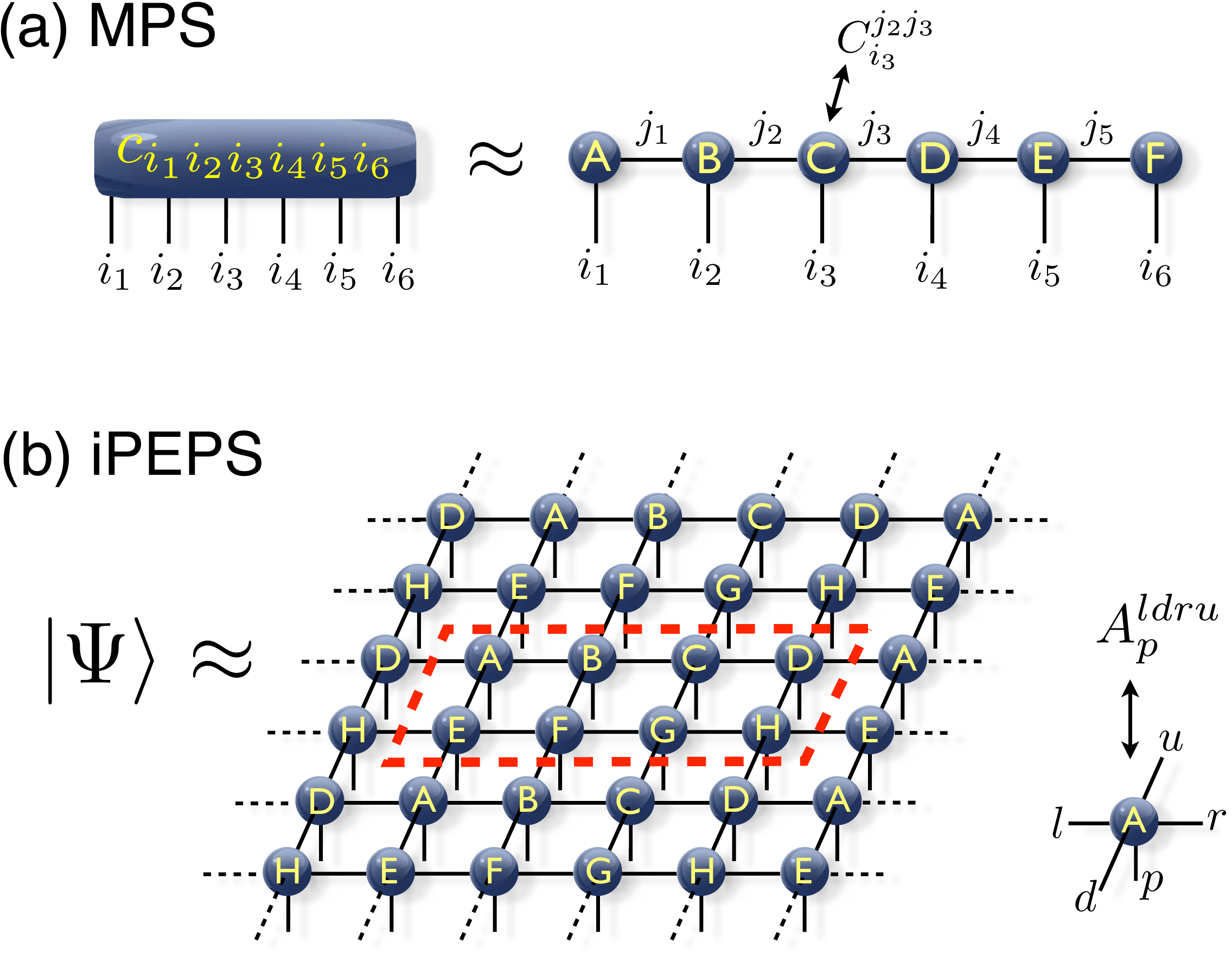}
\caption{(Color online) Graphical representation of an infinite projected entangled-pair state (iPEPS) made of a $4\times 2$ unit cell of tensors (surrounded by thick dashed lines) which is periodically repeated. Each sphere corresponds to a rank-5 tensor and the lines (legs) attached to the sphere represent the indices of the tensor, as shown on the right hand side.}
\label{fig:iPEPS}
\end{center}
\end{figure}

The main idea of a tensor network ansatz is to represent (approximate) the coefficients $c_{i_1 i_2 \dots i_N} $of a wave function,
\begin{equation}
|\Psi \rangle = \sum_{i_1 i_2 \dots i_N} c_{i_1 i_2 i_3 \dots i_N} |i_1\rangle \otimes  |i_2\rangle \otimes \ldots \otimes  |i_N\rangle
\end{equation}
by a trace over a product of tensors.  Here each index $i_k$ runs over the $d$ local basis states of a lattice site. The most famous example are matrix product states (MPS) which form the class of variational states underlying the density-matrix renormalization group (DMRG) method.\cite{white1992} In an MPS the coefficients are given by a trace over product of 3-index tensors $T_{i}^{lr}$ (with 2-index tensors at the boundaries), as for example for a 6-site system
\begin{equation}
c_{i_1 i_2 i_3 i_4 i_5 i_6} \approx \sum_{j_1 j_2 j_3 j_4 j_5} A_{i_1}^{j_1} B_{i_2}^{j_1 j_2} C_{i_3}^{j_2 j_3} D_{i_4}^{j_3 j_4} E_{i_5}^{j_4 j_5} F_{i_6}^{j_5}
\end{equation}
Thus, each coefficient $c_{i_1 i_2 i_3 i_4 i_5 i_6}$ is given by a product of matrices (with vectors at the open boundaries), hence the name matrix product state. Tensor networks are most conveniently represented graphically, as shown in Fig.~\ref{fig:iPEPS}(a) for this particular example. Each tensor is represented by a shape with lines (legs) attached to it, which correspond to the indices of the tensor. A connection between two tensors implies a sum over the corresponding index, and an open leg of a tensor corresponds to the physical index for the local Hilbert space of a site. Each auxiliary index $j_k$ runs over $D$ elements, which is called the bond dimension. Thus, $D$ controls the size of the tensors (or matrices), i.e. the number of variational parameters of the ansatz.

A projected entangled-pair state (PEPS)\cite{verstraete2004} is a natural generalization of a matrix product state to two dimensions. Instead of a three-index tensor, a five-index tensor $T_{i}^{ldru}$ is introduced for each lattice site on a two-dimensional (square) lattice, where each tensor is connected with its four neighboring tensors via the auxiliary indices $l$, $d$, $r$, $u$, each having a bond dimension $D$. Thus, the number of variational parameters per tensor is $dD^4$.
An infinite PEPS (iPEPS) is an ansatz for a wave function in the thermodynamic limit.\cite{jordan2008} It is made of a unit cell of tensors which is periodically repeated on the infinite lattice, as depicted in Fig.~\ref{fig:iPEPS}(b). If the wave function is translational invariant, the same tensor can be used on each lattice site. If the state breaks translational symmetry, a larger unit cell may be required.\cite{corboz2011} In practice, different unit cell sizes are tested to check, which size leads to the state with lowest variational energy.

An iPEPS with $D=1$ is nothing but a site-vectorized wave function (a product state), parametrized by vectors $T_i$ on each site. With increasing $D$ the iPEPS can represent more and more entangled states, with a scaling of the entanglement with block size which obeys the area law of the entanglement entropy.\cite{Verstraete08,eisert2010} Or in other words, with increasing $D$ the iPEPS can take into account more of the quantum fluctuations of the true ground state. These quantum fluctuations may select, e.g. one of infinitely many degenerate states in the classical $D=1$ case. Thus, iPEPS provides a way to systematically study a state as a function of $D$, where $D$ controls the amount of quantum fluctuations (or entanglement) in the system.

In order to obtain an approximate representation of the ground state for a given Hamiltonian, the tensors need to be \emph{optimized}, i.e. the best variational parameters have to be found. In this work we do this optimization by performing an imaginary time evolution of an initial (e.g. random) iPEPS. The evolution operator $\exp(-\beta \hat H)$ is split into a product of two-body operators via a (second order) Trotter-Suzuki decomposition (see e.g. Ref.~\onlinecite{corboz2010}). Application of such a two-body operator to a bond in the iPEPS increases the dimension of the bond. Thus, for an efficient evolution, the corresponding bond needs to be truncated back to the original bond dimension $D$ after each time step. There are different schemes to perform this truncation. In the present work we use the so-called full update, which is explained in detail in Ref.~\onlinecite{corboz2010} for Hamiltonians with nearest-neighbor interactions. In this truncation scheme, the full wave function is taken into account to determine the relevant subspace, in contrast to the \emph{simple update},\cite{vidal2003-1,jiang2008,corboz2010} which is computationally cheaper, but less accurate. In Ref.~\onlinecite{corboz2010-nn} a simple update scheme  for Hamiltonians with next-nearest neighbor interactions was presented. In appendix~\ref{sec:nnscheme} we introduce a more accurate full update scheme to treat next-nearest neighbor interactions.

Once we have obtained an approximate representation of the ground state in form of an iPEPS we can compute expectation values of observables $\hat O$. To do this we have to contract (evaluate) the tensor network representing $\langle \Psi | \hat O | \Psi \rangle$. In an MPS this can be done in an exact way by performing a sequence of pairwise multiplications of tensors. However, in iPEPS (and PEPS) the contraction cannot be done exactly (in polynomial time), and there exist different schemes to perform the contraction approximately. Here we use the corner-transfer matrix method~\cite{nishino1996, orus2009-1} generalized to arbitrary unit cell sizes.~\cite{corboz2011} The accuracy of the approximate contraction is controlled by the "boundary" dimension $\chi$, which we typically choose between 20 up to several hundreds, depending on $D$. For the SSM, this is sufficiently large so that the error due to $\chi$ is small (compared to the effect of the finite $D$). 

\begin{figure}
\begin{center}
\includegraphics[width=8.5cm]{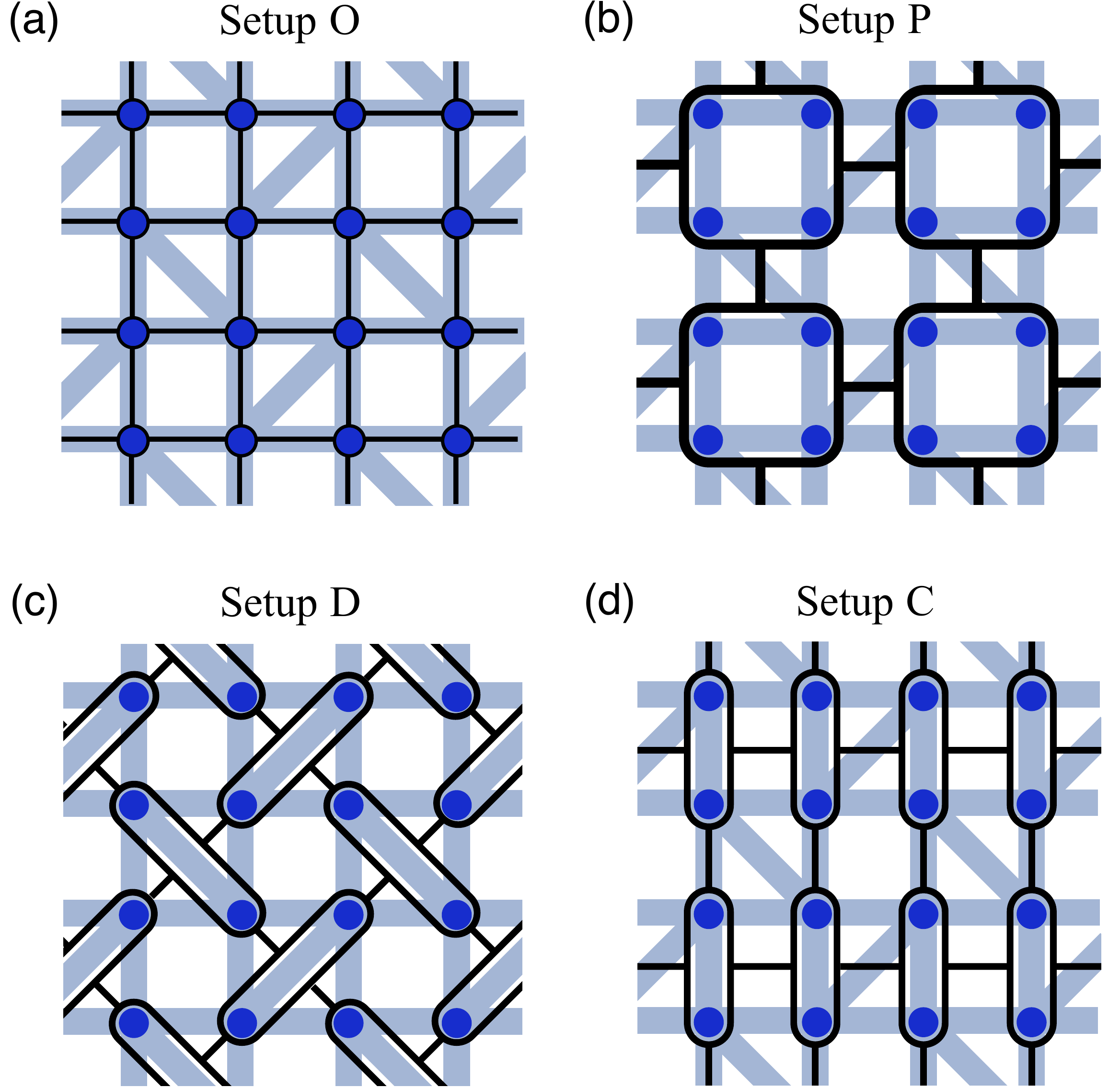}
\caption{(Color online) The four different simulation setups used to simulate the Shastry-Sutherland model with a square lattice iPEPS. Dark shapes correspond to tensors and dark lines to auxiliary bonds between tensors (the open, physical index of each tensor is omitted). Filled circles correspond to physical lattice sites and interactions between the physical sites are given by thick shaded lines. In the different simulation setups different sites are blocked together. A tensor is used for each of these block of sites:
(a) Setup O: one tensor per physical lattice site (no blocking), (b) setup P: one tensor per four sites on a plaquette, (c) setup D: one tensor per orthogonal dimer (two sites), (d) setup C: one tensor per dimer (two sites), arranged in columnar order.}
\label{fig:sssetups}
\end{center}
\end{figure}

\subsection{Order parameters}
Since iPEPS describes a wave function in the thermodynamic limit symmetries such as SU(2) spin rotation or lattice symmetries may be spontaneously broken. To check for SU(2) breaking we compute the local magnetic moment on each site,
\begin{equation}
m=\frac{1}{2} \sqrt{\langle \sigma_x \rangle^2 + \langle \sigma_x \rangle^2  +\langle \sigma_y \rangle^2 }.
\end{equation}
A finite $m$ implies that the SU(2) symmetry is spontaneously broken.

To check for lattice symmetry breaking we measure the local bond energies $E_b$ on each bond in the unit cell, and then compute the following order parameter:
\begin{equation}
\Delta E= \max(E_b) - \min(E_b).
\end{equation}
If $\Delta E$ is finite, then the state breaks lattice symmetry. By plotting the distribution of the different $E_b$ in the unit cell, as e.g. in Fig.~\ref{fig:PD}, we can distinguish between plaquette-, dimer- or other possible orders.

We note that broken symmetries can also be an artifact of a too small $D$ in iPEPS. It is therefore always important to check, how an order parameter behaves as a function of $D$. If it tends to a finite value in the $D\rightarrow \infty$ limit, then there is a true, physical symmetry breaking (long-range order). If, however,  an order parameter is strongly suppressed with increasing $D$ and extrapolates to zero in the infinite $D$ limit, then this indicates that the exact ground state does \emph{not} break the symmetry.

\subsection{Simulation setups}
An iPEPS is a very general ansatz in the sense that the Hamiltonian is essentially the only physical input of a simulation (besides the choice of the unit cell size). The method is to a large extent unbiased, except that simulations for low bond dimensions are biased towards low-entanglement (mean-field) solutions, but this bias disappears for sufficiently large  $D$. Nevertheless, in some cases it can be  useful to bias the state towards a certain order, particularly in order to \emph{rule out} a particular order. For example, if we bias the states towards a dimer state, we are likely to observe a dimer state for small $D$. But if this order vanishes with increasing $D$, then this indicates that the order is unstable.

We can put a bias  by choosing different bond dimensions on the individual bonds in the iPEPS. For example, taking a much larger bond dimension between sites on a plaquette leads to a bias towards a plaquette state. Instead of using different bond dimensions we can also block a certain number of sites together, and use one tensor for each block of sites. This effectively corresponds to having an infinite $D$ between the sites within a block, i.e. all correlations between the sites in a block are taken into account exactly. We note that, no matter how we block the sites, in the large $D$ limit one should always recover the same physics. Thus, trying different setups provides a way to check the robustness of a result. The same idea was used in Ref.~\onlinecite{Corboz12_simplex}, where we explained in detail how to construct the Hamiltonian for the block sites.

In Fig.~\ref{fig:sssetups} the four simulation setups used in this work are presented. Setup O is corresponds to the "original" lattice with one tensor per physical lattice site. The horizontal and vertical shaded lines correspond to nearest-neighbor interactions, and the diagonal shaded lines to next-nearest neighbor interactions. Since there is  no blocking and all bond dimensions are the same, there is essentially no bias in this setup.

Setup P is biased towards a plaquette order by blocking four sites on a plaquette together, leading to a local Hilbert space of $2^4=16$ for each block site. One can see that there are only interactions connecting nearest-neighbor tensors (dark rounded squares). 

In setup D we use a tensor for two sites on each of the orthogonal dimers, which leads to a square lattice iPEPS tilted by 45 degrees, with a local dimension of $2^2=4$ for each block site. This setup is biased towards the orthogonal dimer state. We note that also in this case only nearest-neighbor interactions between the block sites appear.

Finally, in setup C we also group two sites on dimers together, but in a columnar-dimer arrangement, which puts a bias towards columnar-dimer order. The diagonal interactions oriented from bottom left to top right appear between nearest-neighbor tensors, whereas the other diagonal interactions connect next-nearest neighbor tensors.

In setups C and O we make use of the next-nearest neighbor scheme introduced in appendix~\ref{sec:nnscheme}, whereas in the setups P and D only a nearest-neighbor scheme for the interactions between the block sites is needed.

\subsection{Technical remarks on the iPEPS simulations}
There are different possibilities to choose an initial iPEPS for the imaginary time evolution. In many cases, an iPEPS obtained with the simple update provides a good initial state for simulations with the full update (cf. Ref~\onlinecite{corboz2010}). Typically, for small bond dimensions we evolve several random states and check which one has the lowest variational energy. This state is then used as an initial state for simulations at larger $D$. Close to a phase transition it can be useful in some cases to initialize the iPEPS with a state deep inside a phase (away from the phase transition). For example, for the plaquette phase with setup D, we obtained particularly good results from an initial iPEPS which was obtained from simulations of the SSM with a plaquette bias (stronger interactions on plaquettes).

In the imaginary time evolution the energy decreases as a function of $\beta$. In some cases we observed a slight increase of the energy for large $\beta$, probably because the evolution is only performed in an approximate way. In these cases we take the state with lowest variational energy, for the $\beta$ where the energy is minimal.

Our simulation results are all obtained with a $2\times2$ unit cell. We tested larger unit cell sizes and did not find signs of another low-energy state which requires a larger unit cell.

To improve the efficiency of our simulations we used tensors with $\mathbb{Z}_q$ symmetry (a subgroup of SU(2)). The tensors then acquire a block structure, similarly to a block-diagonal matrix. Details on the implementation of global abelian symmetries can be found in Refs.~\onlinecite{singh2010,bauer2011}.

For the corner-transfer matrix scheme we adopted the method explained in Refs.~\onlinecite{orus2009-1, corboz2011}, but instead of computing isometries based on a singular value decomposition to absorb a column (or a row) of tensors into the boundary (see Refs.~\onlinecite{orus2009-1, corboz2011} for details), we use the projector introduced in Refs.~\onlinecite{Wang11, Huang12}.


\section{Simulation results}
\label{sec:results}
Our main results are summarized in the phase diagram in Fig.~\ref{fig:PD}: iPEPS predicts an intermediate plaquette phase between the dimer phase and the N\'eel phase in the range $0.675(2) < J < 0.765(15)$. In the following we first discuss the properties of the individual phases, and then provide a detailed study of the phase transitions.

\subsection{Dimer phase}
The dimer phase consists of a product of exact singlets along the diagonal bonds in the lattice, i.e. for each value $J$ in the dimer phase the ground state is the same, with an energy of $-3/4$ per dimer, or $E_s=-3/8$ per lattice site.

With setup D this state has a trivial representation with iPEPS, i.e. it can be represented with $D=1$ because the state simply corresponds to a product state of the dimers. For the other simulation setups $D=2$ is required, because a singlet cannot be written as a product state of two sites. 

\subsection{N\'eel phase}
\begin{figure}
\begin{center}
\includegraphics[width=8.5cm]{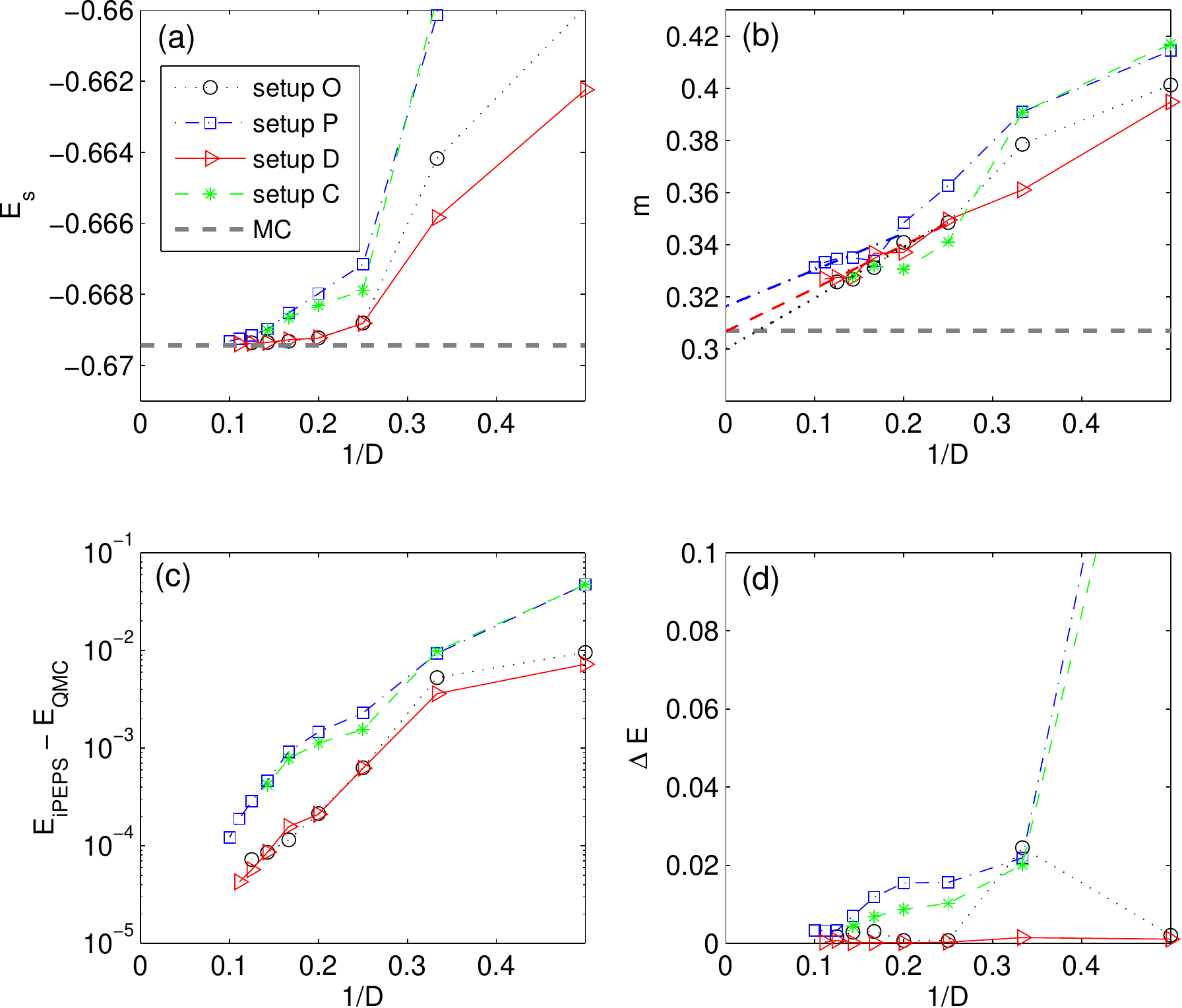}
\caption{(Color online) Benchmark results for the Heisenberg model ($J'=0$, $J=1$) for the different simulation setups, compared with state-of-the-art QMC simulations.  (a) Variational energy as a function of the inverse bond dimension. (b) Local ordered moment $m$ as a function of inverse bond dimension.  The linear extrapolations are a guide to the eye. (c) Deviation of the iPEPS energy from the QMC result. (c) (d) The difference in bond energies $\Delta E$ vanishes in the large $D$ limit for all setups, which shows that lattice symmetries are unbroken.  }
\label{fig:Heisenberg}
\end{center}
\end{figure}

The N\'eel phase in the Shastry-Sutherland model is adiabatically connected to the N\'eel phase of the Heisenberg model, which corresponds to the limit $J\rightarrow \infty$ (or $J'=0$). In this limit the model is no longer frustrated and can therefore be solved by Quantum Monte Carlo (QMC). In Fig.~\ref{fig:Heisenberg} we compare the iPEPS results obtained with the different simulation setups with  state-of-the-art QMC data from Refs.~\onlinecite{sandvik1997,Sandvik10}. Figures~\ref{fig:Heisenberg}(a) and (c) show how the variational energy gets improved with increasing bond dimension $D$. Our best variational energy $E_s^{D=9}=-0.66939$, obtained with setup D, agrees up to four digits with the extrapolated value from QMC, $E_\text{QMC}=-0.669437(5)$.

In all setups the local magnetic moment $m$ decreases with increasing $D$, as shown in Fig.~\ref{fig:Heisenberg}(b), and approaches the QMC value, $m_\text{QMC}=0.30743(1)$.\cite{Sandvik10} It is not known how $m$ depends on the bond dimension $D$ which makes accurate extrapolations to the infinite $D$ limit difficult. Empirically, a linear extrapolation for the largest few values of $D$ yields a reasonable estimate of the value in the infinite $D$ limit, with a relative error of the order of a few percents. This is of course much less accurate than QMC, however, we can obtain results with a similar accuracy also for the cases with finite $J'$, where QMC suffers from the negative sign problem. What really matters in the present study is that we can clearly distinguish between a finite and a vanishing order parameter to identify the different phases, which is clearly feasible with the current accuracy.

Setup P (setup C) is biased towards a plaquette (columnar dimer) state, and this is why for small $D$ one finds a finite value of the plaquette (columnar dimer) order parameter as shown in Fig.~\ref{fig:Heisenberg}(d). However, the plaquette (dimer) order is strongly suppressed with increasing $D$ and vanishes for large $D$. Thus, even though individual setups exhibit a bias for small $D$, eventually they all become exact in the large $D$ limit.

\begin{figure}
\begin{center}
\includegraphics[width=8.5cm]{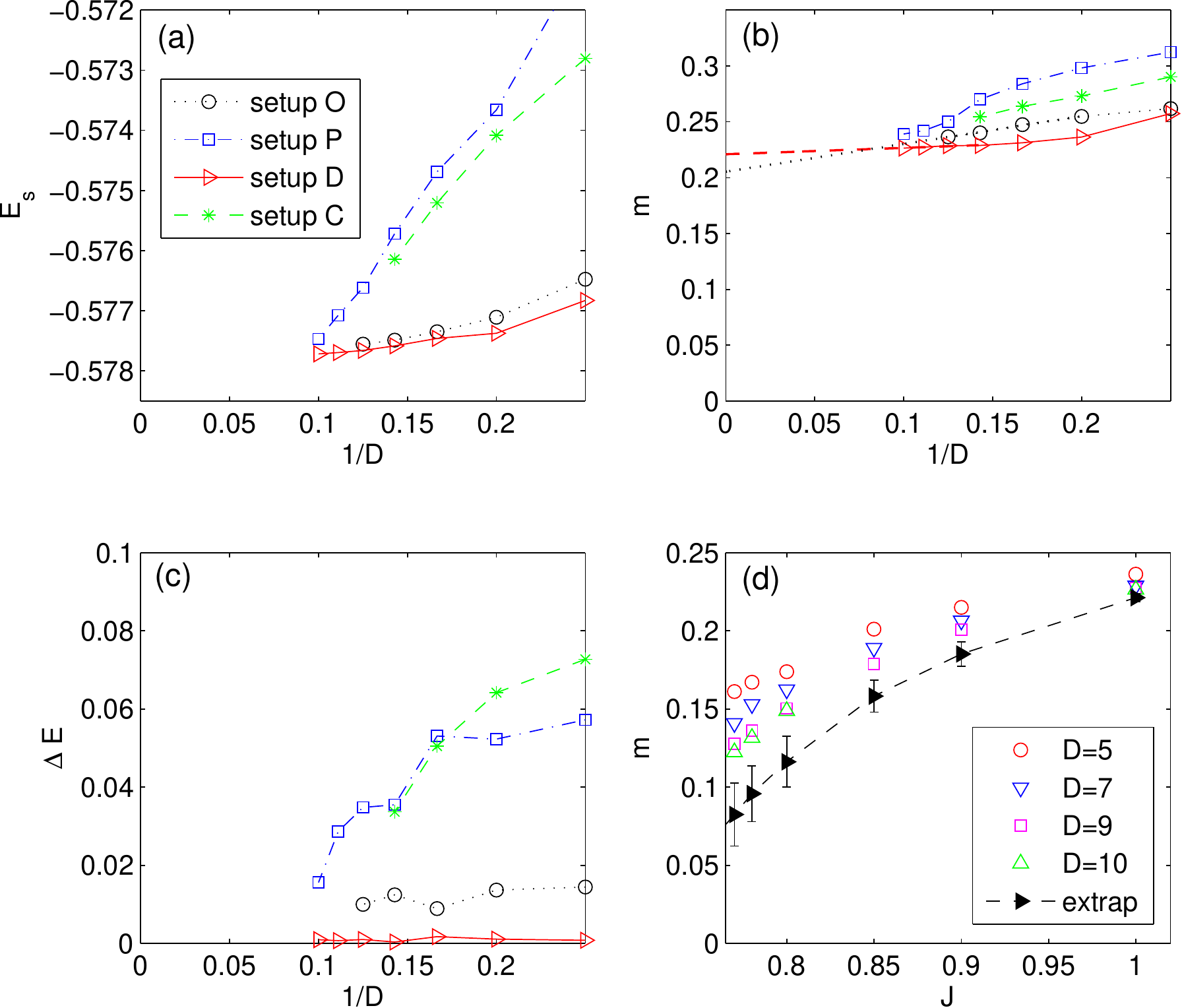}
\caption{(Color online) Results for the Shastry-Sutherland model in the N\'eel phase. (a)-(c) Energy, local order moment $m$, and difference in bond energies $\Delta E$ as a function of inverse bond dimension, obtained for $J=1$ with the four simulation setups.  The linear extrapolations are a guide to the eye. (d) Local ordered moment as a function of $J$ in the N\'eel phase, obtained with setup~D.}
\label{fig:res_ssJ1}
\end{center}
\end{figure}

For finite $J'$ the system is frustrated because the spins on a diagonal bond would like to be antiparallel to each other (instead of parallel as in the N\'eel phase). This competition of interaction leads to a suppression of the local magnetic moment $m$. For example, for $J=1$ (Fig.~\ref{fig:res_ssJ1})  the local moment is reduced to $m=0.21(1)$, and it is further suppressed with decreasing $J$. The difference in bond energies $\Delta E$ also vanishes in this case in the large $D$ limit, indicating the absence of translational symmetry breaking.

\subsection{Plaquette phase}
\begin{figure}
\begin{center}
\includegraphics[width=8.5cm]{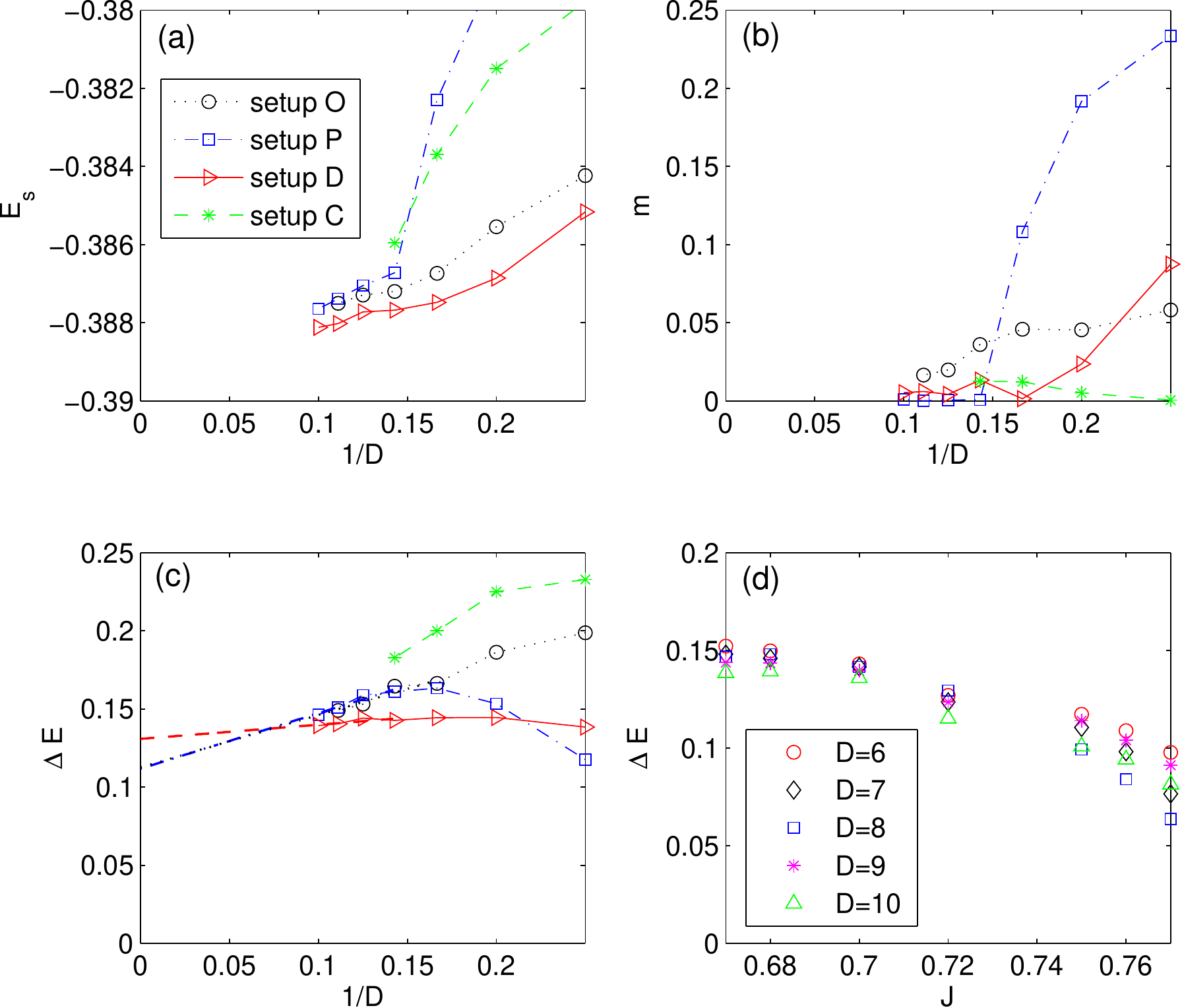}
\caption{(Color online) Results for the Shastry-Sutherland model in the plaquette phase. (a)-(c) Energy, local order moment $m$, and difference in bond energies $\Delta E$ as a function of inverse bond dimension, obtained for $J=0.7$ with the four simulation setups.  The linear extrapolations are a guide to the eye. (d) Plaquette order parameter as a function of $J$ in the plaquette phase obtained with setup~D.}
\label{fig:res_ssJ07}
\end{center}
\end{figure}

In Fig.~\ref{fig:res_ssJ07} we present simulation results for $J=0.7$, where all  setups consistently predict a plaquette phase for large bond dimension. The lowest variational energy obtained is $E_s^{D=10}=-0.3881$ with setup D.

The difference in bond energies $\Delta E$ in Fig.~\ref{fig:res_ssJ07}(c) clearly remains finite in all simulation setups, and  we find that the strong bonds form plaquettes as illustrated in Fig.~\ref{fig:PD} for large $D$. The dependence of $\Delta E$ on $D$ varies from one setup to another, nevertheless, a rather large value between $0.1$ and $0.15$ seems to be compatible with all simulation setups.

The local magnetic moment in Fig.~\ref{fig:res_ssJ07}(b) is finite for small $D$, which would suggest coexistence of plaquette and N\'eel order. However, $m$ is clearly suppressed with increasing $D$ and it is very likely to vanish in all simulation setups in the large $D$ limit, which shows that the SU(2) spin rotation symmetry is not broken in the plaquette phase, as expected.

Finally, in Fig.~\ref{fig:res_ssJ07}(d) the plaquette order parameter as a function of $J$ for different bond dimensions $D$ is shown. Its magnitude tends to decrease with increasing $J$ but remains finite in the whole plaquette phase.

\subsection{Absence of a columnar-dimer phase}
\begin{figure}
\begin{center}
\includegraphics[width=8cm]{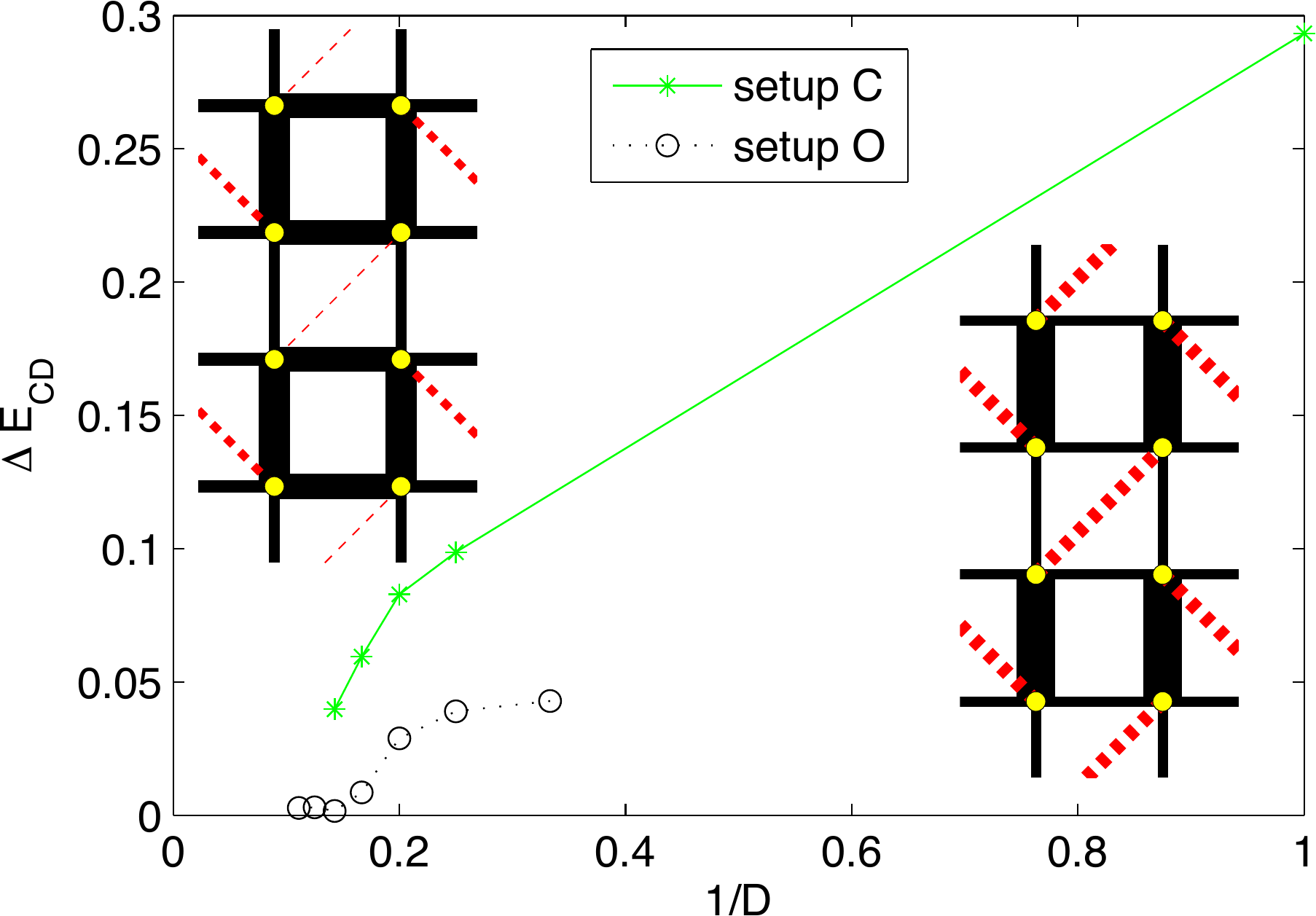}
\caption{(Color online) Columnar-dimer order parameter as a function of inverse bond dimension for $J=0.7$. The insets show the bond energies for simulation setup C, for $D=1$ (right side) and $D=5$ (left side), where the width of a bond is proportional to the magnitude of its energy with full (dashed) lines corresponding to negative (positive) energies. A finite dimer order parameter can be found for small $D$. With increasing $D$, however, the dimer order becomes weaker, and vanishes in the large $D$ limit. }
\label{fig:res_ssdimer}
\end{center}
\end{figure}

In Ref.~\onlinecite{Zheng01} a columnar-dimer state was predicted as one of the possible candidates for an intermediate phase from series expansion calculations. With iPEPS we can also reproduce such a dimer state with setup C which is biased towards this type of order. For $D=1$ a product of vertical dimers is obtained in this case, as shown in the right inset in Fig.~\ref{fig:res_ssdimer}. Upon increasing $D$, however, this dimer order is strongly suppressed with increasing $D$, and plaquette order is enhanced, i.e. two parallel horizontal bonds connecting two dimers become stronger, as shown in the left inset in Fig.~\ref{fig:res_ssdimer}. We define the following columnar-dimer order parameter
\begin{equation}
\Delta E_{CD} = \max(E_{bx}) - \max(E_{by})
\end{equation}
with $E_{bx}$ and $E_{by}$ the bond energies in horizontal and vertical direction, respectively. This order parameter can be finite for small $D$, but even for setup C, it is seen to vanish in the infinite $D$ limit, as shown in Fig.~\ref{fig:res_ssdimer}. Thus, the dimer state can appear as a low-entanglement solution, but eventually quantum fluctuations destroy the dimer order and give rise to a plaquette order instead.

\subsection{First order phase transition between the dimer phase and the plaquette phase}
In this section we determine the transition point $J_{c1}$ between the dimer phase and the plaquette phase. The transition is clearly of first order, as can be seen in the jump in the plaquette order parameter at $J_{c1}$ from zero to a finite value.

To find the transition value precisely we can make use of the hysteresis effect in the vicinity of a first order phase transition: When we initialize a state in the dimer phase and tune $J$ to a value slightly above $J_{c1}$ the state will remain in the dimer phase (since the state is metastable). Similarly, a state initialized in the plaquette phase will remain in that phase when decreasing $J$ to a value slightly below $J_{c1}$. The phase transition occurs where the energies of the two states cross. Since the energy in the dimer phase is independent of $J$, $E_s=-0.375$, the transition occurs when the energy of the plaquette state crosses $E_s=-0.375$.

\begin{figure}
\begin{center}
\includegraphics[width=8cm]{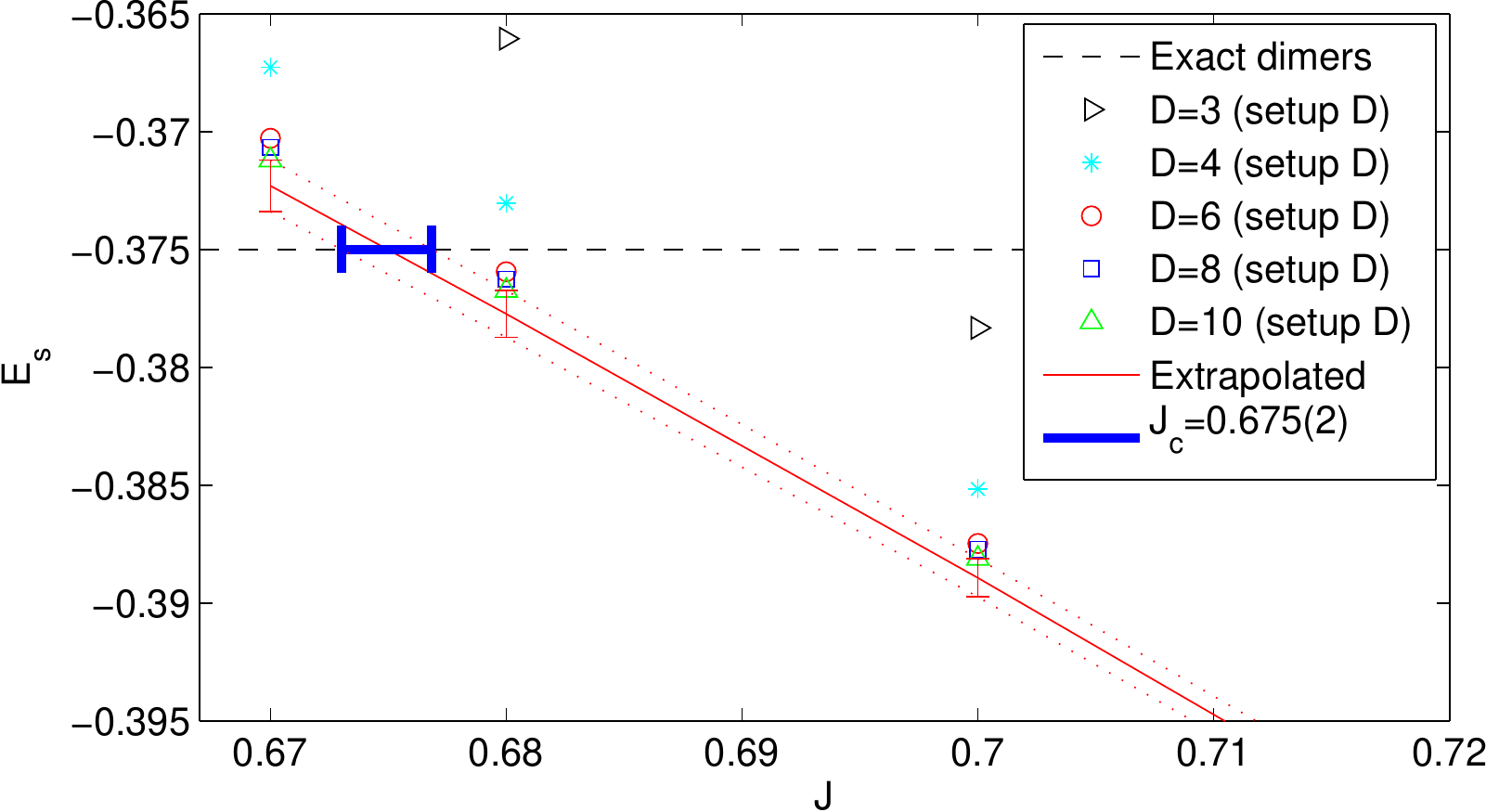}
\caption{(Color online) Energy of the dimer state (horizontal dashed line) and the extrapolated energy of the plaquette state (full line). The first order phase transition occurs at $J=0.675(2)$ where the two energies cross. }
\label{fig:res_ss_first}
\end{center}
\end{figure}

In Fig.~\ref{fig:res_ss_first}, the energy of the plaquette state as a function of $J$  is shown, obtained with setup D which yields lowest variational energies. Since with increasing $D$ the energy of the plaquette state is decreasing, the crossing point $J_{c1}^D$ obtained for a certain $D$ is an upper bound for $J_{c1}$. For example, for $D=10$ we find $J_{c1}^{D=10}=0.6768$. If we extrapolate the energy linearly in $1/D$ using the three largest values of D, we obtain $J_{c1}^{D\rightarrow \infty}=0.6730$, which we take as an estimate for the lower bound, since the energy is seen to converge faster than linear in $1/D$. Thus, we find for the transition value $J_{c1}=0.675(2)$, which  is in agreement with the series expansion result $0.677(2)$ from Ref.~\onlinecite{Koga00}, and the prediction from a higher-order coupled cluster method ($0.677$).\cite{Darradi05}

\subsection{First-order phase transition between plaquette phase and N\'eel phase}
Locating the phase transition between plaquette phase and the N\'eel phase is the most challenging part of this work. We find that the transition is of (weak) first order.
As discussed in the previous section the transition occurs where the energies of the two states cross. However, unlike in the transition between the dimer and plaquette phase, the energies cross at a very small angle, which makes it difficult to very accurately determine the crossing point.

\begin{figure}
\begin{center}
\includegraphics[width=8.5cm]{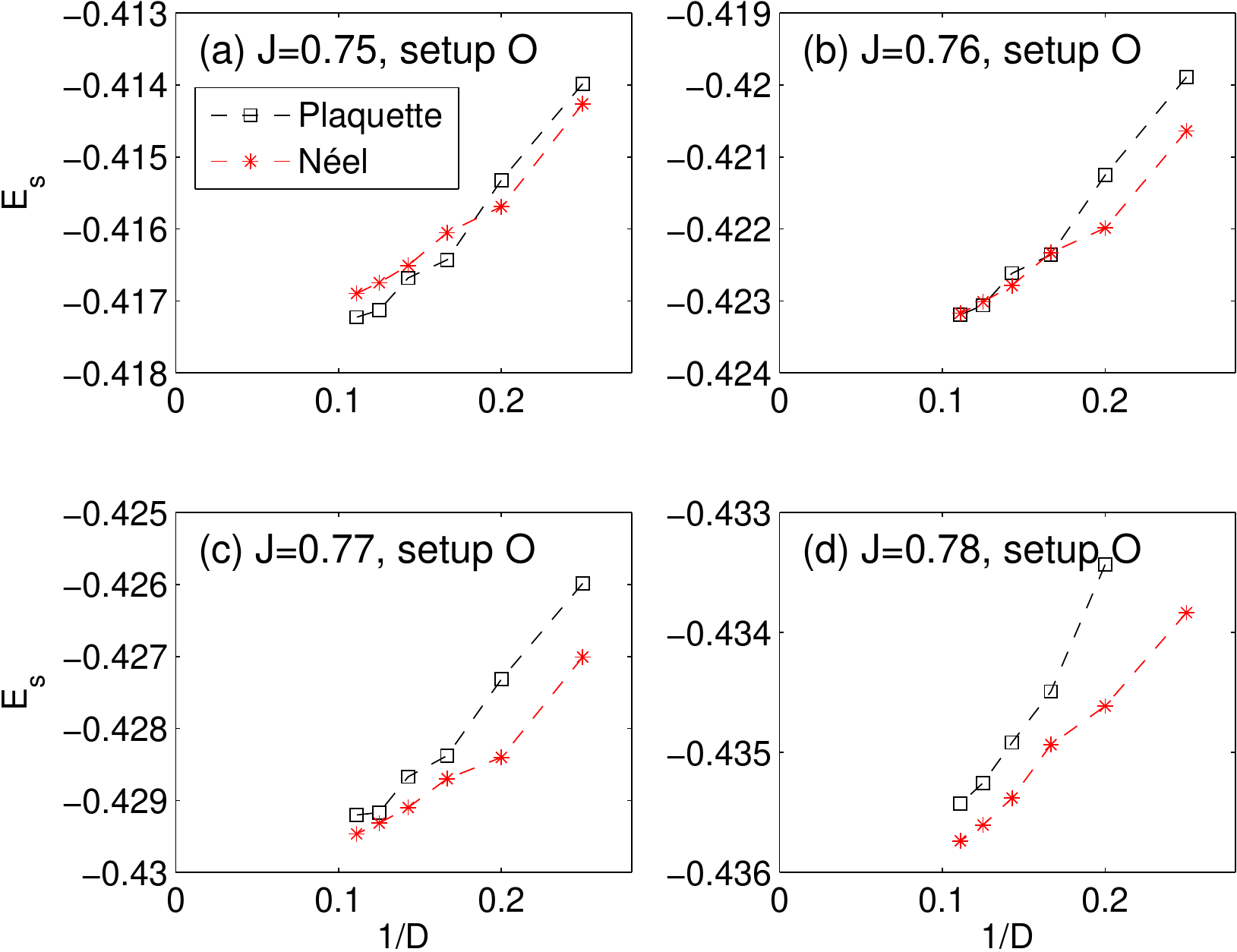}
\caption{(Color online) Energies of the plaquette state and the N\'eel state in the vicinity of the first order phase transition between the two states, obtained with setup O. For $J=0.76$ the two states have a similar energy  for large $D$. For $J=0.75$ ($J=0.77$) the plaquette state has a lower (higher) energy than the N\'eel state.  }
\label{fig:res_ss3e}
\end{center}
\end{figure}

In Fig.~\ref{fig:res_ss3e} we plot the energies of the two states in the vicinity of the transition as a function of $1/D$ obtained with setup O. For $J=0.75$ the plaquette state has a lower energy than the N\'eel state for large bond dimensions. By increasing $J$ the energy of the N\'eel state is lowered with respect to the plaquette state, so that for $J=0.76$ they become essentially equal at large bond dimensions, and finally for $J=0.78$ the N\'eel state has clearly a lower energy than the plaquette state. Since the two curves have similar slopes we do not expect another crossing of the energies at larger $D$.

\begin{figure}
\begin{center}
\includegraphics[width=8.5cm]{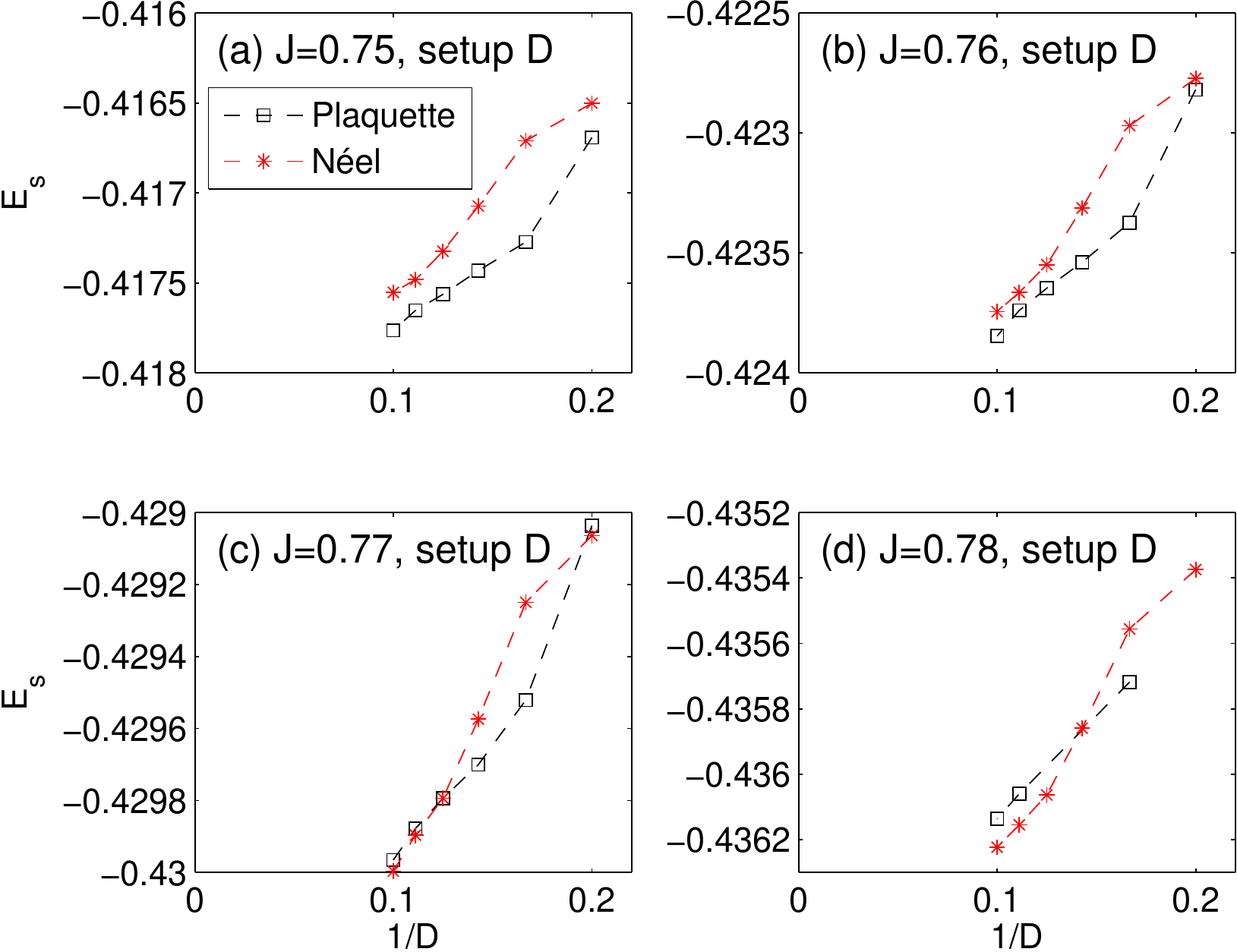}
\caption{(Color online) 
Energies of the plaquette state and the N\'eel state in the vicinity of the first order phase transition between the two states, obtained with setup D. For $J=0.77$ the two states have a similar energy  for large $D$. For $J=0.76$ ($J=0.78$) the plaquette state has a lower (higher) energy than the N\'eel state. }
\label{fig:res_ss3eD}
\end{center}
\end{figure}

A similar result is obtained with setup D, shown in Fig. \ref{fig:res_ss3eD}, where the two energies become similar for $J=0.77$ for large $D$.
Thus, these results suggest a first order phase transition occurring for $J_{c2}=0.765(15)$ which is much smaller than the predicted value $0.86(1)$ from series expansion.~\cite{Koga00}

\section{Conclusion}
\label{sec:conclusion}
In this work we showed that the intermediate phase in the Shastry-Sutherland model is a plaquette phase, in agreement with several previous studies based on other methods.\cite{Koga00,Takushima01,Laeuchli02} Using state-of-the-art iPEPS simulations we have accurately determined the phase boundaries of the plaquette phase. By using different simulation setups we biased the solution towards different ground states. But despite the bias, all setups consistently predict an intermediate plaquette phase, and we found that the previously suggested columnar-dimer phase is unstable. 

This work provides a good basis for future studies of the SSM in a finite magnetic field, a subject which is still attracting a lot of attention.  Indeed, despite a considerable theoretical effort,\cite{Miyahara99, Momoi00,Dorier08,  Abendschein08, Nemec12} the actual sequence of plateaus of the plain SSM has not been definitely established yet, at least close to the transition to the plaquette  phase, and to get reliable information on the magnetization curve in this parameter range is an important step towards the interpretation of the numerous and partially conflicting experimental results reported for SrCu$_2$(BO$_3$)$_2$. \cite{Kageyama99, Onizuka00, Kodama02,Sebastian08,Takigawa12}

Our study further demonstrates the usefulness of iPEPS as a tool to study strongly correlated systems in 2D. By simulating a model directly in the thermodynamic limit we can minimize possible boundary effects, which is an advantage over DMRG, where typically long cylinders with a certain width $W$ are simulated. For small widths DMRG is clearly more accurate than PEPS, however, since the number of relevant states to be kept in DMRG scales exponentially with $W$, DMRG becomes less accurate than PEPS for systems exceeding a certain width. For the Heisenberg model this seems to be the case around $W\sim 10$, \cite{stoudenmire2011} where the accuracy in the energy for $W=10$ and $m=3000$ states is comparable to the accuracy of an iPEPS with $D=9$ for the system in the thermodynamic limit (for a  PEPS with a finite width $W=10$ and $D=9$ the accuracy would presumably be better). Thus, the two approaches can be seen as complementary: DMRG provides very accurate results up to a certain system size, and iPEPS approaches the problem from a different limit: Instead of varying the system size the only essential parameter is $D$, which controls the amount of entanglement in the system. While in many cases it is understood how to perform finite size extrapolations, little is known on how to extrapolate quantities in $D$. Obtaining a better understanding on how quantities depend on $D$ will be important to determine order parameters more accurately in future studies.

Finally, it is remarkable to note that in the above mentioned example for the Heisenberg model the total number of variational parameters is roughly three orders of magnitude larger in DMRG (MPS) than in iPEPS, which shows that a PEPS offers a much more compact description of a 2D wave function than an MPS. The challenge, however, remains to further improve the methods to optimize and contract an (i)PEPS, so that even larger bond dimensions and larger accuracies can be achieved. A lot of progress has been made in the last few years, however, we believe that a method which fully exploits the potential of the ansatz is yet to come.

{\it Note added. ---} After completion of this work, we became aware of a new study of the SSM based on the multi-scale entanglement renormalization (MERA) ansatz in Ref.~\onlinecite{Lou12}, in which conclusions similar to ours have been reached for zero magnetic field. In particular, both methods locate the transition between the plaquette and the N\'eel phase
significantly below the estimate of Ref.~\onlinecite{Koga00}. The critical value for the phase transition between the orthogonal dimer phase and the plaquette phase reported in that paper, $J=0.687(3)$, is also not far from our result, but it lies between our $D=3$ and $D=4$ estimates, hence definitely above the upper bound $J_{c1}^{D=10}=0.6768$ provided by iPEPS. This indicates that the MERA result is closer to the mean-field solution than the iPEPS result.


%
\acknowledgments
The simulations have been performed on the Brutus cluster at ETH Zurich. We thank the support of the Swiss National Science Foundation and MaNEP.

\appendix

\section{Nearest-neighbor full update}
\label{sec:nnscheme}
In this section we describe the full update used to perform an imaginary time evolution with a Hamiltonian including next-nearest neighbor interactions. The description is intended for readers who are familiar with the basic notions of tensor network algorithms. For a general introduction we refer to Refs.~\onlinecite{Verstraete08,schollwoeck2011}. The idea of the following scheme is essentially the same as for the nearest-neighbor full update discussed in Ref.~\onlinecite{corboz2010}: We apply a local time evolution operator $\hat U=\exp(-\tau \hat H_l)$ to the iPEPS that represents a wave function $|\Psi \rangle$, to obtain an evolved wave function (in imaginary time),
\begin{equation}
|\Psi' \rangle = \hat U_l |\Psi \rangle
\end{equation}
This increases the bond dimension between the sites, on which $\hat H_l$ is acting on, from $D$ to some $D'=\kappa D$. The crucial step is now to truncate the enlarged bond indices from $D'$ back to $D$, which yields an approximate wave function $|\tilde \Psi\rangle$ of $|\Psi' \rangle$. To do this we minimize the norm of the distance between the two wave functions,
\begin{equation}
\label{eq:normdist}
 ||  |\Psi' \rangle - |\tilde \Psi\rangle  ||^2 =   \langle \Psi' |\Psi' \rangle - \langle \Psi' |\tilde \Psi \rangle - \langle \tilde \Psi | \Psi' \rangle + \langle \tilde \Psi | \tilde \Psi \rangle
\end{equation}
with respect to the parameters of the new tensors in $| \tilde \Psi \rangle$.

\begin{figure}[htb]
\begin{center}
\includegraphics[width=8.5cm]{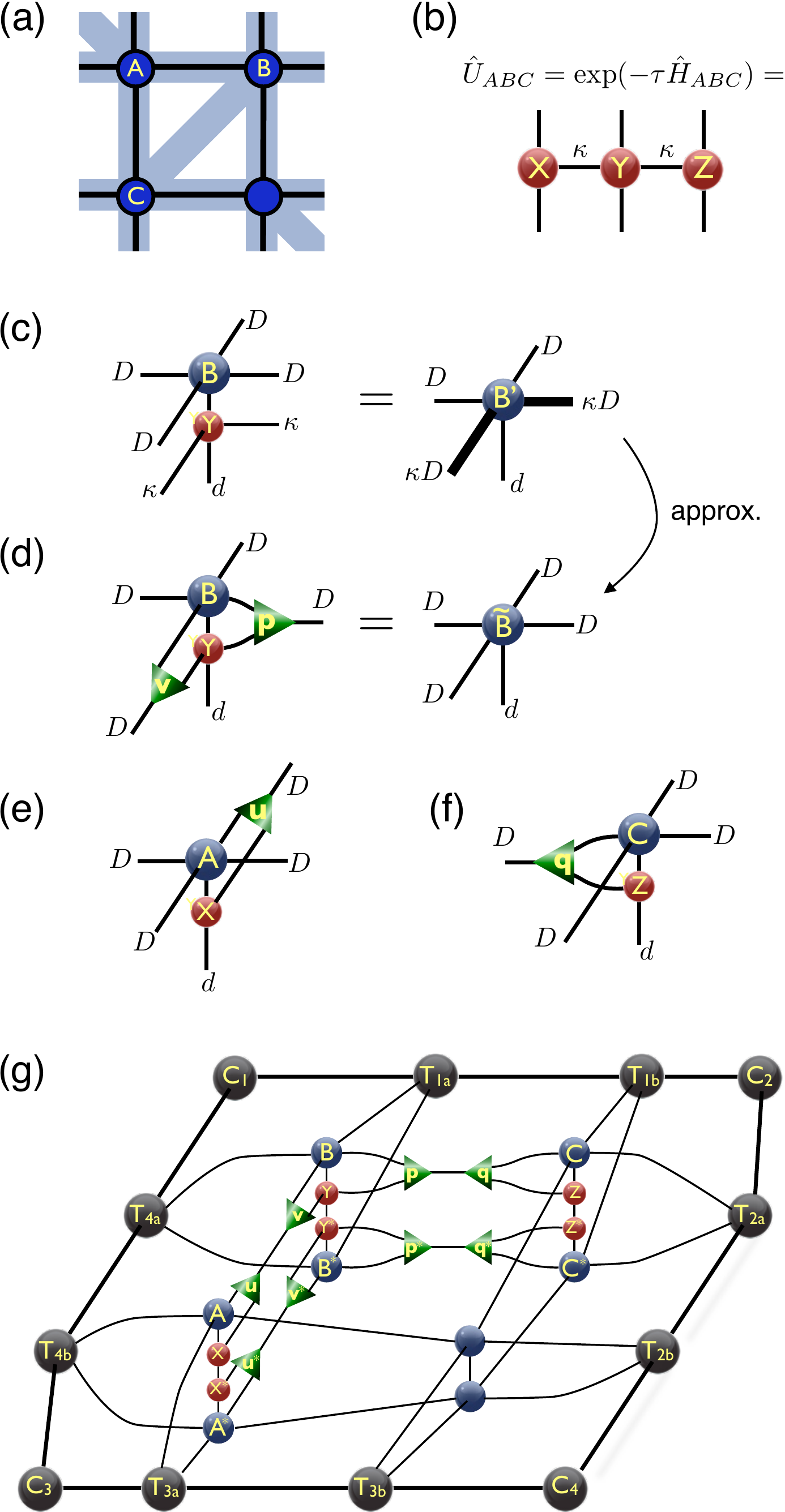}
\caption{(Color online) (a) We consider three sites A,B,C on a triangle in the Shastry-Sutherland model (cf. Fig.~\ref{fig:sssetups}). (b) The imaginary time evolution operator on the triangle represented as a matrix product operator (MPO). (c) MPO tensor $Y$ multiplied to the PEPS tensor $B$. This product can be represented by another PEPS tensor $B'$ with enlarged bond dimensions $\kappa D$ in the lower and right leg. (d) Ansatz for the new tensor $\tilde B$ to approximately represent $B'$. (e-f) Similar ansatz as in (d) for $\tilde A$ and $\tilde C$. (g) Tensor network representation of $\langle \tilde \Psi | \tilde \Psi \rangle$ (see text).  }
\label{fig:nnnupdate}
\end{center}
\end{figure}

As an example, lets consider the interactions on the triangle made of the sites A,B and C in Fig.~\ref{fig:nnnupdate}(a). The SSM has three two-body terms on this triangle, which we denote by $\hat H_{\text{ABC}}$. The corresponding imaginary time evolution operator $\hat U_{\text{ABC}}=\exp(-\tau \hat H_{\text{ABC}})$ can be exactly represented by a matrix product operator (MPO) with a certain bond dimension $\kappa$, as shown in Fig.~\ref{fig:nnnupdate}(b), with tensors $X$, $Y$, and $Z$ acting on sites A, B, C, respectively. By multiplying these MPO tensors to the corresponding PEPS tensors, e.g. $Y$ to $B$ as in Fig.~\ref{fig:nnnupdate}(c), we obtain a new PEPS tensor $B'$ with increased bond dimension $\kappa D$ on the lower and the right leg. The aim is now to approximate tensor $B'$ with another tensor $\tilde B$ with all bond dimensions reduced to $D$. For this we make an ansatz for tensor $\tilde B$ as shown in Fig.~\ref{fig:nnnupdate}(d), made of the original tensor $B$ and $Y$, and two tensors $v$ and $p$ of dimension $D\times \kappa \times D$. Similarly, we make an ansatz for $\tilde A$ and $\tilde C$ as shown in Figs.~\ref{fig:nnnupdate}(e) and (f), respectively.

With these ingredients we can represent the wave functions $|\Psi \rangle$, $|\Psi' \rangle$, $|\tilde \Psi \rangle$, as iPEPSs, where the new wave function depends on the parameters of the tensors $u$, $v$, $p$, and $q$. In order to obtain the optimal new iPEPS we need to minimize
\begin{equation}
\label{eq:normdist2}
F(u,v,p,q) = \langle \tilde \Psi_{(u,v,p,q)} | \tilde \Psi_{(u,v,p,q)} \rangle  - 2 \text{Re}( \langle  \tilde \Psi_{(u,v,p,q)} |  \Psi'  \rangle )
\end{equation}
where we omitted the constant term $\langle \Psi' |\Psi' \rangle$ in \eqref{eq:normdist}. The tensor network for the first term is shown in Fig.~\ref{fig:nnnupdate}(g). As usual, the surrounding corner tensors $C_k$ and edge tensors $T_j$ are obtained from the corner-transfer matrix method.\cite{nishino1996, orus2009-1,corboz2011}  They take into account the infinite wave function (times its conjugate) surrounding a $2\times2$ cell of tensors. The representation of the second term looks similar, except that the tensors $u$, $v$, $p$, $q$  are replaced by identities (straight lines).

As in the case of the nearest-neighbor update~\cite{corboz2009} we can minimize this sum either iteratively, or e.g. with a conjugate gradient method. In this work we used an iterative scheme, where expression \eqref{eq:normdist2} is first minimized with respect to $u^*$ with the other tensors being fixed, i.e.
\begin{equation}
\label{eq:normdist2}
 \frac{\delta}{\delta u^*}  F(u,v,p,q)  = \frac{\delta}{\delta u^*} \left[ u^* M u - u^* K \right] = 0
\end{equation}
where $M$ is the tensor network in Fig.~\ref{fig:nnnupdate}(g) with $u$ and $u^*$ removed, and K is the tensor network representation of $\langle \tilde \Psi_{(u,v,p,q)} |\Psi'   \rangle$ with the tensor $u^*$ removed. Thus, the minimum can be found by solving the  system of linear equations
\begin{equation}
\label{eq:normdist3}
M u = K,
\end{equation}
which yields a new tensor $u$. One then proceeds with tensor $v$, keeping tensors $u$, $p$, $q$ fixed, and so on. This procedure is repeated until convergence is reached.

\appendix

\bibliographystyle{apsrev4-1}
\bibliography{../bib/refs}

\end{document}